\documentclass[conference]{IEEEtran}
\usepackage{cite}
\usepackage[T1]{fontenc} % optional

\usepackage{amsmath,amssymb,amsfonts}
\usepackage[cmintegrals]{newtxmath}
\usepackage{bm} % optional
\usepackage{threeparttable}
\usepackage{booktabs}
\usepackage{tablefootnote}
\usepackage{algorithm}
\usepackage{algpseudocode}
\usepackage{graphicx}
\usepackage{subcaption}
\usepackage{textcomp}
\usepackage{tikz}
\usetikzlibrary{positioning}
\usepackage{multirow}
\usepackage{listings}
\usepackage[table]{xcolor}
\usepackage{siunitx}
\usepackage{tikz}
\usepackage{pgfplots}
\pgfplotsset{compat=1.18} \usepgfplotslibrary{statistics}
\usepackage{arydshln}

\usepackage{color}
\usepackage{listings}
\usepackage{caption}
\usetikzlibrary{calc}

\newcommand{\blinded}[1]{#1}

\def\BibTeX{{\rm B\kern-.05em{\sc i\kern-.025em b}\kern-.08em
    T\kern-.1667em\lower.7ex\hbox{E}\kern-.125emX}}
\begin{document}

\title{Precomputed 1D-CNNs for Atrial Fibrillation Detection on Tiny Smart Sensor Systems}
%\title{Atrial Fibrillation Detection for Tiny Smart Sensor Systems with Precomputed 1D-CNN Hardware}
\author{
    \IEEEauthorblockA{
    Lukas~Einhaus, Natalie~Maman, Julian~Hoever, Andreas~Erbslöh, Gregor~Schiele\\
    \textit{Intelligent Embedded Systems Lab, University of Duisburg-Essen},\\
    \textit{Faculty of Computer Science, Duisburg, Germany} \\
    \{first-name\}.\{last-name\}@uni-due.de
    }
    % ORCID 
    % Lukas {\textcolor{blue}{\texttt{ORCID: 0000-0002-6102-7077}}}
    % Julian {\textcolor{blue}{\texttt{ORCID: 0009-0001-2906-8645}}}
    % Andreas {\textcolor{blue}{\texttt{ORCID: 0000-0001-6702-892X}}}
    % Gregor {\textcolor{blue}{\texttt{ORCID: 0000-0003-4266-4828}}}
}
\maketitle
\begin{abstract}
1D-CNNs play a crucial role for time-series analysis on tiny smart sensor systems, e.g. for biosignal analysis, predictive maintenance, or structural health monitoring. 
LUT-based precomputation has emerged as an interesting optimization technique to implement such neural networks on FPGAs. 
The core idea is to precompute all possible outputs of a neural network layer and store them directly in the lookup tables of the FPGAs. 
This enables highly resource-efficient networks with ultra-low latency but suffers from poor scalability.
Previous work has explored using depthwise-separable convolutions to improve scalability. 
In this paper, we generalize this approach to consider additional forms of grouped convolutions. 
Based on this, we propose a novel type of convolutional block and an algorithm to guide the choice of hyper parameters for this block.
We evaluate our approach on a medical time-series dataset for predicting atrial fibrillation using the MIT-BIH database~(ECG~recordings).
% Comparing our approach with depthwise-separable convolutions, we achieve a similar accuracy while reducing the consumption of FPGA resources by a factor of~2 down to~1,662 with a F1-Score of~94\% for detecting atrial fibrillation. 
The resulting hardware accelerators are small enough to be deployed on an AMD~Spartan-7~S15. They achieve a F1-score of up to 95.6\% while only requiring 2,844 LUTs and no DSPs or BRAM.

\end{abstract}
\newcommand{\onedcnn}{1D-CNN}
\newcommand{\torch}{PyTorch}
\newcommand{\lut}{LUT}
\newcommand{\lutbased}{LUT-based}
%%%
\section{Introduction}
Smart sensor systems combine sensing technology with on-device machine learning to process sensor data locally. This enhances privacy, reaction time, and offline availability.

One dimensional convolutional neural networks~({\onedcnn}s) have been shown to be a well suited machine learning approach for this, usable in a diverse set of application fields, such as structural health monitoring~\cite{avci2017structural}, predictive maintenance~\cite{ince2016real, eren2019generic} and various medical applications~\cite{kiranyaz2015real, li2017classification, mattioli20221d, Loehler2024}.
While transformer architectures gained a lot of traction in recent years, 1D-CNNs can still provide a more suitable trade-off between performance and resource requirements. Especially in very small sensor systems, transformer architectures can quickly exceed time or energy constraints~\cite{ling2025_strikewatch}.
To speed up the computation of~{\onedcnn}s, our work uses small-scale Field Programmable Gate Arrays (FPGAs) like the  Spartan 7 S15 to implement efficient hardware accelerators~\cite{mittal2020survey}. FPGAs contain thousands of lookup tables (LUTs) that can implement small Boolean functions, e.g., a function with 4 or 6 inputs (each 1 bit) and an 1 bit output, depending on the selected FPGA. Together, these can emulate arbitrary hardware circuits, ranging, e.g., from adders and multipliers to whole CPUs. 

Our work goes one step further, making use of the fact that LUTs can reduce arbitrarily complex functions to a value lookup of constant time complexity.
Using this, we \textit{precompute} neural network components, such as a filter in a convolutional layer or a neuron in a linear layer and represent them as truth tables that can be stored in the FPGA's LUTs.
This is especially useful for ultra-low latency applications. 
The main challenge with this approach is to limit the exponential growth of the precomputed truth tables, which depends on the number of input bits to the implemented function.
We will refer to this number as fan~in.
For a function with~$I$ inputs with a bitwidth of~$b$, the resulting truth table is made up of~$2^{b{\cdot}I}$ entries. 
Therefore, it is crucial to minimize fan~in for each network component as much as possible. 
Previous work has resorted to random sparsity~\cite{umurogluLogicNetsCoDesignedNeural2020, andronic2023polylut} or lossy function representations for dense neural networks~\cite{nazemiNullaNetTinyUltralowlatency2021}.

For~{\onedcnn}s, previous work has used depthwise separable convolutions to reduce the fan~in for precomputed convolutions~\cite{einhausPrecomputed1DConvolutionalLayers2021}. This works well for small kernels but does not scale for more complex architectures.
Therefore, it is necessary to explore more possibilities for replacing conventional convolutions. In this paper, we propose to use the more general concept of \textit{grouped convolutions}~\cite{krizhevsky2012_imagenet} instead of depthwise separable convolutions. This can lead to extreme cost reductions while achieving similar accuracy.
More specifically, we make the following contributions:
\begin{itemize}
    \item We propose a configurable grouped convolutional block to replace conventional dense convolutions. Such blocks can be used to split a convolution into a sequence of two convolutions that can be precomputed and represented with a small number of LUTs on FPGAs.
    \item We introduce a score to rate the above-mentioned configurations wrt. network performance vs. resource consumption to help selecting favourable configurations without having to try out all possible configurations.
    %and evaluate our approach on atrial fibrillation detection~\cite{moody1983new}.
    \item We provide an open source tool to generate synthesize-able hardware designs based on our approach and evaluate their performance on real hardware.
\end{itemize}

The remainder of our paper is structured as follows. After a discussion of related work in Section~\ref{sec:relWork}, we present our approach for precomputing~{\onedcnn}s on FPGAs in Section~\ref{sec:approach}. We evaluate our approach in Section~\ref{sec:eval} using the MIT-BIH dataset~\cite{moody1983new} and conclude the paper in Section~\ref{sec:conclusion}.

\section{Related Work}\label{sec:relWork}
%%%% 
In the following section, we discuss related work for LUT-based precomputation approaches in deep neural networks.

A number of approaches tries to reduce the fan~in of network components. 
LogicNets~\cite{umurogluLogicNetsCoDesignedNeural2020} shows an approach for networks composed of linear layers.
The authors reduce area consumption by initializing neurons with random sparsity.
PolyLUT~\cite{andronic2023polylut} presents a generalization of LogicNets, replacing all computations between the input and output of a neuron by a single~LUT. They consider not only linear transformations of the input data, but also polynomial ones, thus taking advantage of the fact that a LUT can represent arbitrary Boolean functions. The resulting model has a higher flexibility to fit the data, potentially needing fewer layers to represent the same function. 
The increase in LUT~size is thus compensated by the layer savings which at the same time improves latency.
% The authors of NullaNet~\cite{nazemiNullaNetTinyUltralowlatency2021} control the size of precomputed truth-tables by restricting the number of values that need to be represented between layers. This is done by setting all values not observed during training to "don't care". The authors report savings in area utilization of up to~\SI{20}{\%}.
Similarly, NeuraLUT~\cite{andronic2024neuralut} represents whole sub-networks by a single~LUT. These sub-networks are composed of multiple fully-connected layers that use floating-point precision inside sub-nets, while keeping the cost low by using quantization and sparsity between sub-nets.

Another set of approaches focusses on compressing truth tables of precomputed functions. Khatei et al.~\cite{khatei2024_compressedLUT} decompose truth tables and combine the results during runtime using simple primitives.
Others additionally record which input values can actually be observed during training and inject \emph{don't care} for everything else to increase the potential for compression~\cite{ebrahimiIterativePruningAlgorithm2023, nazemiNullaNetTinyUltralowlatency2021, cassidy2025_reducedLUT}.

Petersen et al. make networks composed of primitive binary logic gates~(LGNs) differentiable, leading to ultra-sparse low complexity neural networks \cite{petersen2022_LogicGateNetwork} and apply that approach to 2D-CNNs maintaining~\SI{80.17}{\%} accuracy on CIFAR-10, while decreasing inference times by a factor of~160~\cite{petersen2024_LogicGateConv}. 

Weightless neural networks~(WNNs)~\cite{aleksander2009_introWNN} use \emph{RAM-Nodes} (essentially truth tables) as building blocks for neural networks also during training.
Consequently, they are typically not trained with gradient descent.
Bacellar et al. extend the approach for differentiable logic gates from Petersen and apply it to WNNs to obtain differentiable WNNs~(DWNs), improving accuracy compared to NeuraLUT, while reducing resource consumption roughly by a factor of up to~200.
Nag et al. present a vision transformer~(ViT) architecture incorporating DWNs~\cite{Nag_2025}. While this improves accuracy from~\SI{57.5}{\%} to~\SI{95.5}{\%} on CIFAR-10 compared to Bacellar et al., it drastically increases the number of LUTs from~46k to~587k.

None of these approaches focuses on 1D-CNNs for ultra low-power FPGAs with only a few thousand LUTs. 

\section{Our Approach} \label{sec:approach}
%Next, we describe our tool workflow to train a neural network and to translate it into hardware. Afterwards, we discuss the steps for deploying grouped convolutions using precomputed 1D-CNNs with binarizations.
We propose an automated way to derive LUT-based hardware accelerators for resource-constrained smart sensor systems from existing off-the-shelve 1D-CNNs defined in PyTorch.
To achieve this, our approach has to decrease the fan~in of the network's components drastically and explore the trade-off between resource consumption and network performance.
The obtained network architectures are trained as usual and automatically translated to a hardware description language (HDL).
The resulting accelerators do not need any block~RAM~(BRAM) or DSP slices for multiplications. 

Our approach can be broken down into five steps: (i) binarize activations, (ii) use our Split Convolutional Block, that is based on grouped convolutions, to replace dense convolutions, (iii) filter the number of potential candidates for said replacements heuristically, (iv) identify precomputable structures and compute their corresponding truth tables, (v) convert the result into HDL.

\subsection{Binary Activations for Precomputable Blocks}
Replacing full-resolution with binary activations is not only a straight-forward way to reduce the fan~in, it also allows for fine-grained control of the fan~in by varying the input channels of convolutions.
We binarize activations by using
%%%
\begin{align}
    \text{bin}(x) = \begin{cases} 1 &\text{if } x \geq 0 \\ -1 &\text{else}\end{cases}
\end{align}
%%%
in the forward pass and replace the gradient with the straight-through-estimator~(STE)~\(\frac{\partial}{\partial x} \text{bin} = 1\) for the backward pass~\cite{courbariauxBinarizedNeuralNetworks2016}.
Weights will only be represented implicitly on hardware, and are kept at full resolution.

We use the binary activations in the network architecture to identify blocks that are precomputable, i.e., that can be represented via LUTs. 
We can include more components between two binary activations to increase the capacity of precomputed blocks without additional cost, as long as this does not increase the fan~in of the corresponding block.
%%%%%%%%%%%%%%%%%%%%%%%%%%%%%%%%%%%%%%%%%%%%%%%%%%%%%%
\subsection{LUT Cost} \label{sec:lut_cost}
Before describing our Split Convolutional Block, we first introduce an analytical cost model to estimate the required physical LUT count. This model will also be used by our toolchain, later. Without it, we would need to synthesize costs from HDL, which takes a lot of time.

As this analytical model does not include any heuristic logic optimizations, it represents a worst case estimation.
The actual resource consumption of hardware resulting from synthesis will typically be much lower.
%%%
\begin{figure}[htpb]
%   \begin{subfigure}{0.95\linewidth}
  \centering
  \scalebox{0.85} { % scalebox as scaling in tikz does alter distance between nodes
\begin{tikzpicture}[scale=1, every node/.style={font=\small}]
  % First level: two 3-to-1 gates
  \node[draw, rectangle, minimum width=1cm, minimum height=1cm] (G1) at (0,1.3) {};
  \node[draw, rectangle, minimum width=1cm, minimum height=1cm] (G2) at (0,0) {};
  % Second level: 2-to-1 gate
  \node[draw, rectangle, minimum width=1cm, minimum height=1cm] (G3) at (2.2,0) {};
  % Inputs
  \node[left=0.5 of G1] (I01) {$X_1$};
  \node[left=0.5 of G1, yshift=0.4cm] (I00) {$X_0$};
  \node[left=0.5 of G1, yshift=-0.4cm] (I02) {$X_2$};
  \node[left=0.5 of G2] (I11) {$X_1$};
  \node[left=0.5 of G2, yshift=0.4cm] (I10) {$X_0$};
  \node[left=0.5 of G2, yshift=-0.4cm] (I12) {$X_2$};
  % Connect I0, I1, I2 to both G1 and G2
  \draw[->] (I00) -- ++(0.4,0) |- ($(G1.west) + (0, 0.2)$);
  \draw[->] (I01) -- ++(0.4,0) |- ($(G1.west) + (0, 0)$);
  \draw[->] (I02) -- ++(0.4,0) |- ($(G1.west) + (0, -0.2)$);
  \draw[->] (I10) -- ++(0.4,0) |- ($(G2.west) + (0, 0.2)$);
  \draw[->] (I11) -- ++(0.4,0) |- ($(G2.west) + (0, 0)$);
  \draw[->] (I12) -- ++(0.4,0) |- ($(G2.west) + (0, -0.2)$);
  % Connect I3 to G3 as select
  \node[below=0cm of I12] (I3) {$X_3$};
  \draw[->] (I3) --  ++(2.2, 0) |- ($(G3.west) + (0, -0.2)$);
  % Outputs from G1 and G2 to G3
  \draw[->] (G1.east) -- ++(0.4,0) |- ($(G3.west) + (0, 0.2)$);
  \draw[->] (G2.east) -- ++(0.4,0) |- (G3.west);
  % Output
  \node[right=1.1cm of G3] (Y) {$Y$};
  \draw[->] (G3.east) -- (Y);
\end{tikzpicture}}
\caption{Exemplary composition of higher fan~in ($n=4$) truth table composed of three 3:1-LUTs}
\label{fig:lut_composition}
\vspace*{-1mm}
\end{figure}
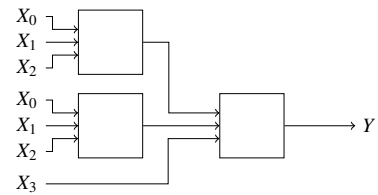
%%%
Similarly to the pattern shown in Figure~\ref{fig:lut_composition}, one can construct~\(n\):1-truth tables using~3:1-LUTs in hardware. We denote the LUT count for a~\(n\)~bit truth table with~\(\mathcal{C}_n\).
Using AMD FPGAs like Spartan~7 with 6:1-LUTs, there we distinguish between three cases:~\(n \leq 6\), \(n = 7 + 2l\) (for odd cases) and~\(n = 8 + 2l\) (for even cases) with~\(l \in \mathbb{N}_0\). 

For~\(n \leq 6\) we can directly implement the truth table with a single LUT. 
For~\(n = 7 + 2l\), we obtain
%%%
\begin{equation}
  \mathcal{C}_n = 2 \cdot \mathcal{C}_{n-1} + 1.
\end{equation}
%%%
For~\(n = 8 + 2l\), we can build the truth tables from four instances of the LUT tree for~\(n-2\) and combine the four results with an additional~LUT leading to
\begin{equation}
  \mathcal{C}_n = 4 \cdot \mathcal{C}_{n-2} + 1 = 2 \cdot (\mathcal{C}_{n-1} - 1) + 1.
\end{equation}
Thus, we can estimate the cost of composing a~$n$-to-1~truth table from~LUTs using the following recursive relation:
\begin{align}
\mathcal{C}_n &= \begin{cases}
  1 &\text{if } n \leq 6 \\
  2 \cdot \mathcal{C}_{n-1} - (-1)^n &\text{else} \label{eq:cn_recursive}
\end{cases}
\end{align}

From that, we derive the closed form expression for the LUT cost for implementing a~\(X\)-to-\(Y\)~truth table (required from logical function) using~6:1-LUTs by:
\begin{equation}
    \mathcal{C}(X, Y) = \frac{Y}{3} \cdot \left(2^{(X-4)} - (-1)^X\right) \label{eq:lutcost}
\end{equation}
We will use this to estimate the cost of convolutional blocks and the resulting neural network.
%%%
\subsection{Split Convolutional Blocks}
\label{sec:split_convs}
Binary activations alone are usually not enough to obtain hardware accelerators deployable on small FPGAs. 
Therefore we additionally replace convolutions with our \emph{Split Convolutional Blocks}, which use grouped convolutions to reduce fan~in and thus improve scalability. 

Grouped convolutions generalize the concept of depthwise convolutions.
While a depthwise convolution partitions~\(c\) channels into~\(g\) groups with~\(c=g\) and processes each of them separately, we can choose the number of groups freely for grouped convolutions, as long as 
\begin{equation}
    \exists s_{\text{in}}, s_{\text{out}}\in\mathbb{N} \colon s_{\text{in}}\cdot g = c \land s_{\text{out}}\cdot g = f,
\end{equation}
where \(f\) is the number of output channels and \(s_{\text{in}/\text{out}}\) denote the sizes of input and output groups respectively.

\begin{figure}[htpb]
\centering
% \begin{subfigure}{0.45\linewidth}
%     \centering
%     \rotatebox[origin=c]{270}{\includegraphics[height=\textwidth]{images/DenseConv.pdf}}
%     \caption{Dense convolution with \((c_0, k_0, g_0, f_0) = (4, 2, 1, 2)\) \label{fig:dense_conv}}
% \end{subfigure}
% \hspace*{1em}
\begin{subfigure}{\linewidth}
    \centering
    \includegraphics[height=2.7cm, width=0.75\textwidth, trim={0 0.4cm 0 0.3cm}, clip]{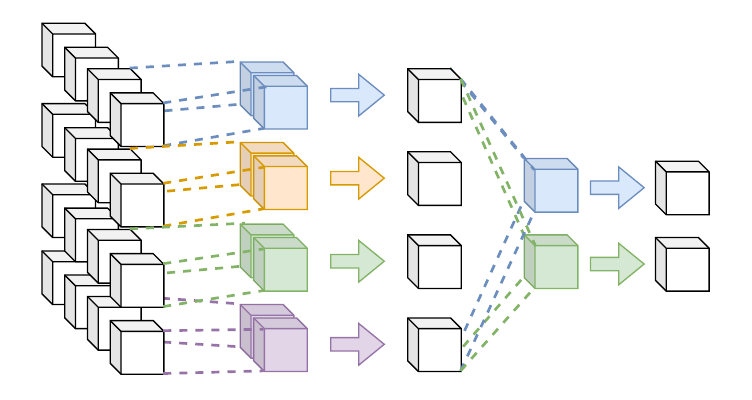}
    \caption{Depthwise separable convolution with \((c_\alpha, k_\alpha, g_\alpha, f_\alpha) = (4, 2, 4, 4)\) and \((c_\beta, k_\beta, g_\beta, f_\beta) = (4, 1, 1, 2)\) \label{fig:depthwise_sep_conv}}
\end{subfigure}
\vspace*{1em}
\begin{subfigure}{\linewidth}    
    \centering
    \includegraphics[height=2.7cm, width=0.75\textwidth, trim={0 0.4cm 0 0.3cm}, clip]{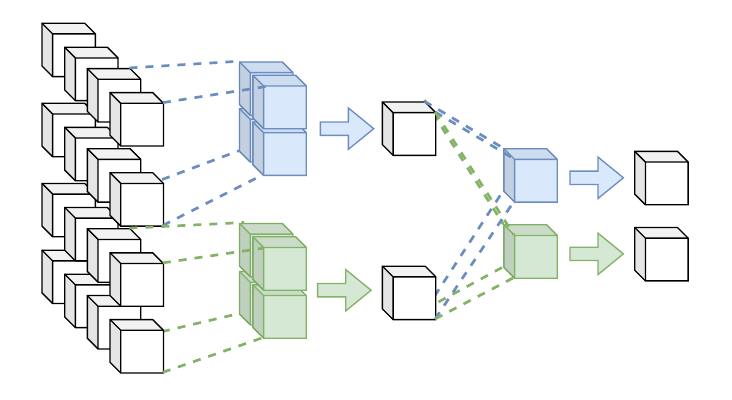}
    \caption{Grouped convolutional block with \((c_\alpha, k_\alpha, g_\alpha, f_\alpha) = (4, 2, 2, 2)\) and \((c_\beta, k_\beta, g_\beta, f_\beta) = (2, 1, 1, 2)\) \label{fig:grouped_conv_block}}
\end{subfigure}

\caption{Visualisation of two different convolutional kernels with different colours. We denote the hyper parameter of the convolutional blocks~\(F = (c, k, g, f)\), with~\(c\) input channels, a kernel size~\(k\), number of groups~\(g\) and~\(f\) output channels.}
\label{fig:three_types_conv}
\vspace*{-1.8em}
\end{figure}

Figure~\ref{fig:three_types_conv} shows in (a) a depthwise separable convolution and in (b) a grouped convolutional block with hyper parameters that allow them to act as a structural replacement for an original dense convolution with four input channels, two output channels and a kernel size of two.
%%%
%%%%
This means, that both the number of input channels~\(c\) and the number of output channels~\(f\) have to be dividable by~\(g\).
With \(k\) denoting the kernel size, we want to replace a convolution with parameters~\(F_0 = (c_0, k_0, g_0, f_0)\) by two convolutions with parameters~\(F_i = (c_i, k_i, g_i, f_i), \, i\in \{\alpha, \beta\}\) and binary activations.
This requires the following condition 
\begin{equation}
\begin{split}
    c_0 = c_\alpha = s_{\text{in}, \alpha} \cdot g_\alpha \; \land\; &f_\alpha = c_\beta = s_{\text{out},\alpha}\cdot g_\alpha = s_{\text{in}, \beta} \cdot g_\beta \\
    \land\; & f_0 = f_\beta = s_{\text{out},\beta} \cdot g_\beta. \\
\end{split}
\label{eq:split_condition}
\end{equation}
We restrict~\((k_\alpha, k_\beta) =  (k_0, 1)\), i.e., a Split Convolutional Block ends with a pointwise convolution, similar to depthwise separable convolutions.

Thus, our proposed Split Convolutional Block (see Figure~\ref{fig:maxpool_reorder}, green area) consists of a convolution with hyper parameters~\(F_\alpha\) followed by a batch normalization, a binary activation function and another convolution with hyper parameters~\(F_\beta\), to replace the original dense convolution with hyper parameters~\(F_0\), such that~\(F_0, F_\alpha, F_\beta\) fulfill the condition from Eq.~\eqref{eq:split_condition}.
The resulting LUT cost can be computed as 
\begin{equation}
    \mathcal{C}\left(k_\alpha \cdot \frac{c_0}{g_\alpha}, f_\alpha\right) + \mathcal{C}\left(k_\beta \cdot \frac{f_\alpha}{g_\beta}, f_0\right)
    \label{eq:split_cost}
\end{equation}
This provides fine-grained control of the fan in and hence the LUT cost by varying \(g\).
%%%%%%%%%%%%%%%%%%%%%%%%%%%%%%%%%%%%%%%%%%%%%%%%%%%%%%%%%%%
\subsection{Treatment of Pooling Layers}
\label{sec:ordering}
% \begin{figure}[htpb]
% \centering
% \begin{minipage}[t]{.4\linewidth}
%   \centering
%   \includegraphics[width=0.7\linewidth]{images/training_block_structure-2.pdf}
%   \subcaption{}
%   \label{fig:training_block}
% \end{minipage}
% \hfill
% \begin{minipage}[t]{.4\linewidth}
%   \centering
%   \includegraphics[width=0.7\linewidth]{images/on_device_block_structure.pdf}
%   \subcaption{}
%   \label{fig:on_device_block}
% \end{minipage}
% \caption{Comparison of convolutional block ordering during training (a) and on-device deployment (b).}
% \label{fig:block_structures}
% \end{figure}
%%%%
%%%
In XNOR-Net~\cite{rastegariXNORnetImagenetClassification2016}, the authors argue that applying a max pooling layer after a binary activation results in significant information loss and thus hurts network performance.
They propose a block structure of conv-pool-bnorm-binarize to replace a structure of conv-bnorm-binarize-pool.
But for our approach this would combine a convolution and a pooling layer in a single precomputable block.
This increases the receptive field of that block and consequently the fan~in.

\begin{figure}[htpb]    \centering
    \includegraphics[trim = 0 31 0 31, clip, width=0.9\linewidth]{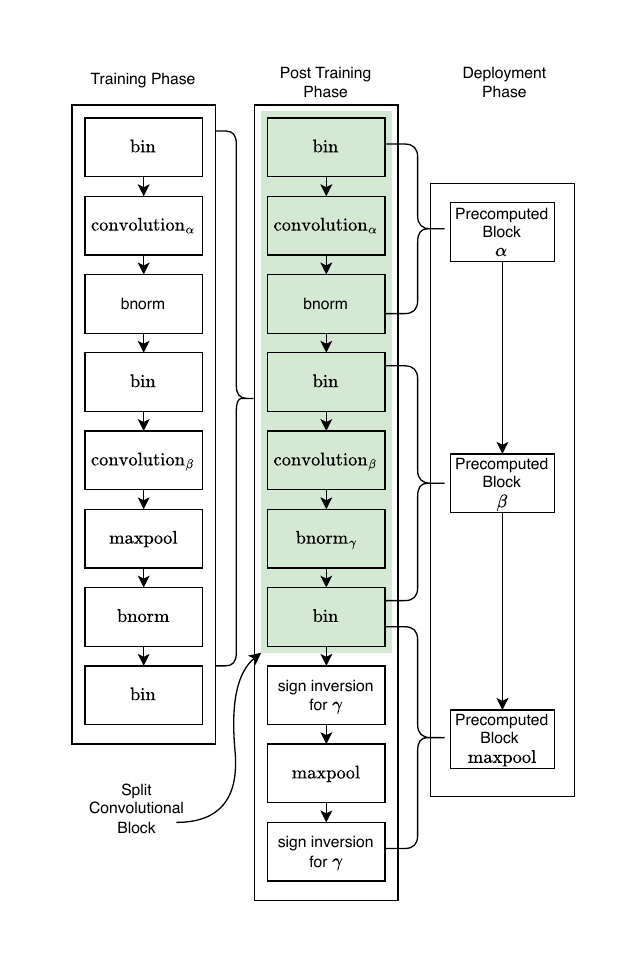}
    \caption{Different stages of the neural network. The Split Convolutional Block is highlighted green.\label{fig:maxpool_reorder}}
\end{figure}

But we can reorder layers for training, such that we use order with higher accuracy during training and our original order with lower resource consumption for precomputation. The left side of Figure~\ref{fig:maxpool_reorder} shows the change of layer order and the added sign inversion operation after convolution \(\beta\) between training and post training phase.
% Moving the max pooling layer between the convolution and the binarization means that the precomputable block now includes not only convolution and batchnorm, but also the max pooling layer.
% Therefore, we have to take both (i) the window size of the max pooling layer and (ii) the stride of the preceding convolutional layer into account.
%As an example we consider a pooling window of size 2. In the worst case, the convolution has a stride that equals its kernel size. Two steps of the convolution are required to produce enough data to fill the pooling window. Therefore the fan in for the corresponding component has doubled.
% More specifically, assuming a pooling window of size~\(p\) and a convolution with stride~\(s\) we find that including the pooling window in the precomputable block will increase the fan~in and result in a total cost of
% \begin{equation}
%     \mathcal{C}\left((\text{min}(s, k)\cdot (p - 1) + k)\cdot \frac{c}{g}, f\right)
% \end{equation}
% We avoid this problem by exploiting the fact that all involved operations are monotonic.
% This allows us to reorder the layers after training, 
%so we can represent the block with a sequence of two logical LUTs instead of one, 
% leading to a cost of
% \begin{equation}
%     \mathcal{C}\left(k \cdot \frac{c}{g}, f\right) + \mathcal{C}(p, f).
% \end{equation}
% Figure~\ref{fig:maxpool_reorder} visualizes the different orders during training and after training, i.e., for precomputation.
To see why and how we can reorder these layers, let
% To avoid the increasing LUT-cost due to ..., we have to make sure that all involved operations are monotonic for reordering the layers. To see this let
\begin{gather}
    \text{bnorm}_\gamma(x) = \frac{x - \mu}{\sigma^2} \cdot \gamma - \beta
\end{gather}
be the batchnorm layer, with~\(\mu \geq 0\), \(\sigma > 0\), \(\gamma \in \mathbb{R}\) and~\(\beta \in \mathbb{R}\).
Let
\begin{equation}
% \begin{split}
    h(x_1, x_2) = \text{bnorm}_\gamma\left(\text{max}(x_1, x_2)\right)
    % &= \frac{\text{max}(x_1, x_2) - \mu}{\sigma^2} \cdot \gamma - \beta.
% \end{split}
\end{equation}
With \(x_1 > x_2\) we have
\begin{equation}
    h(x_1, x_2) = \text{bnorm}_\gamma(x_1) %\frac{x_1 - \mu}{\sigma^2} \cdot \gamma - \beta
\end{equation}
and
\begin{align}
    x_1 > x_2 \Rightarrow \begin{cases}\text{bnorm}_\gamma(x_1) > \text{bnorm}_\gamma(x_2) \text{ if } \gamma > 0 \\ %\frac{x_1 - \mu}{\sigma^2} \cdot \gamma - \beta > \frac{x_2 - \mu}{\sigma^2} \cdot \gamma - \beta \text{ if } \gamma > 0 \\
    \text{bnorm}_\gamma(x_1) < \text{bnorm}_\gamma(x_2) \text{ else}  %(\frac{x_1 - \mu}{\sigma^2} \cdot \gamma - \beta < \frac{x_2 - \mu}{\sigma^2} \cdot \gamma - \beta \text{ else}
    \end{cases}.
\end{align}
Thus, for \(\hat{h}\) with
\begin{equation}
    \hat{h}(x_1, x_2) = \begin{cases}
        \text{max}(\text{bnorm}_\gamma(x_1), \text{bnorm}_\gamma(x_2)) \text{ if } \gamma > 0 \\
        -\text{max}(-\text{bnorm}_\gamma(x_1), -\text{bnorm}_\gamma(x_2)) \text{ else}
    \end{cases}
\end{equation}
it follows that
\begin{equation}
    h(x_1, x_2) = \hat{h}(x_1, x_2).
\end{equation}
Because binarization is monotonous, we can reorder layers by multiplying each channel where the batch norm layer has a~\(\gamma < 0\) with~\(-1\) before and after the pooling layer.
%%%
%%%
While the order of these layers does not impact the computed result, it does change which weights are updated during the backward pass.
Applying pooling first will filter out the features with smaller amplitude.
%%%%%%%%%%%%%%%%%%%%%%%%%%%%%%%%%%%%%%%%%%%%%%%%%%%%%%%
\subsection{Choosing Split Configurations}
Another important design decision is how to choose the split between the two sequential convolutions in our split convolution blocks. 
These blocks provide a large degree of flexibility. 
But they also introduce a complex relationship between hyper parameters.
% The number of channels~\(f_\alpha=c_\beta\) between the convolutions~\(h_\alpha\) and~\(h_\beta\) can be freely chosen.
% But the possible number of groups~\(g\) for each of these layers depends on their input and output channels~\(c, f\).
% The more common divisors each pair~\((c_0, f_\alpha)\), \((f_\alpha, f_0)\) has, the more choices for~\(g_\alpha\) and~\(g_\beta\) are possible.

Choosing a split configuration for each layer independently can quickly result in unmanageable large search spaces.
Therefore, we introduce a score to rate Split Convolutional Blocks prior to training, that is derived from split configurations.
As a first step, we introduce a new metric, we call cross layer connectivity~(CLC), to quantify how many of the input channels~\(c_\alpha\) can influence each of the output channels~\(f_\beta\) of a split convolution.
We express the fan~in of a component as~\(\phi=k\cdot \frac{c}{g}\) and filter out all split convolutions with
\begin{equation}
    \phi_{\text{max}} \geq \text{max}(\phi_\alpha, \phi_\beta) = \text{max}\left(k_{\alpha} \cdot \frac{c_{\alpha}}{g_{\alpha}},\; k_{\beta} \cdot \frac{c_{\beta}}{g_{\beta}}\right).
\end{equation}
We use this constraint in Algorithm~\ref{alg:split} to build a set of possible split configurations~\(\mathcal{F}\).
%%%
\begin{algorithm}[htpb]
\footnotesize
\caption{Finding set of split configurations \(\mathcal{F}\) for a given filter \(F_0 = (k_0, c_0, f_0, g_0)\) and a maximum fan in \(\phi_{\text{max}}\)}\label{alg:split}
\begin{algorithmic}[1]
\Function{FindFilterPairs}{$F_0$}
 \State $seqs \gets \{(k_0, 1), (1, k_0)\}$ \Comment{Possible kernel size sequences}
 \State $\mathcal{F} \gets \emptyset$
 \ForAll{$k_\alpha, k_\beta \in seqs$}
   \State $d_\alpha \gets \emptyset$
   \For{each $g_\alpha$ in divisors of $c_0$}
      \State $\phi_\alpha \gets (c_0 \div g_\alpha) \cdot k_\alpha$
     \Comment{first layer fan in}
     \If{$\phi_\alpha \leq \phi_{\text{max}}$}
       \State $d_\alpha \gets d_\alpha \cup \{g_\alpha\}$
     \EndIf
   \EndFor
   \ForAll{$g_\alpha \in d_\alpha$}
     \ForAll{$g_\beta$ in divisors of $f_0$}
       \State $cs \gets \emptyset$, $c \gets g_\alpha$
       \State $\phi_\beta \gets (c \div g_\beta) \cdot k_\beta$ \Comment{second layer fan in}
       \While{$\phi_\beta \leq \phi_{\text{max}}$}
         \If{$c \mod g_\beta = 0$}
           \State $cs \gets cs \cup c$
         \EndIf
         \State $c \gets c + g_\alpha$
         \State $\phi_\beta \gets (c \div g_\beta) \cdot k_\beta$
       \EndWhile
       \ForAll{$c \in cs$}
         \State $\mathcal{F}\gets \mathcal{F} \cup \{((k_\alpha, c_0, c, g_\alpha),$ 
         \Statex \hspace{\algorithmicindent} $ (k_\beta, c, f_0, g_\beta))\}$
        \EndFor
       \EndFor
     \EndFor
 \EndFor
  \State \Return $\mathcal{F}$
\EndFunction
\end{algorithmic}
\end{algorithm}
%%%
%%%%%%%%%%%%%%%%%%%%%%%%%%%%%%%%%%%%%%%%%%%%%%%%%%%%%%%%%%
\subsubsection{Cross Layer Connectivity (CLC)}
\begin{figure}[htpb]
\centering
\includegraphics[width=0.8\columnwidth, trim={0 0.25cm 0 0.25cm}, clip]{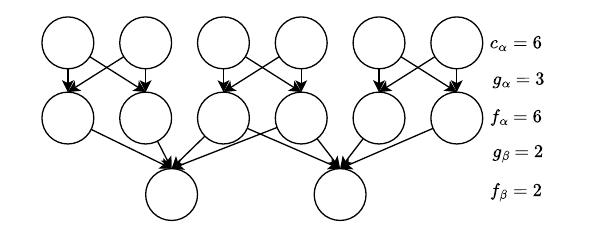}
\caption{Exemplary visualization of a split convolution with a CLC of \(\frac{2}{3}\). Nodes represent an input or output channel. \label{fig:clc_example}}
\end{figure}

So far we have no way to estimate the impact of the resulting split configurations on the network performance.
Previous work showed that connections across groups can have a positive impact on accuracy~\cite{zhang2018shufflenet}.
In the following we want to quantify that connectivity.
We denote the total number of input channels in layer~\(\alpha\) that can influence a single output channel of layer~\(\beta\) by \(c^*\) and  the total number of groups from layer~\(\alpha\), that are processed by each individual filter kernel of layer~\(\beta\) by \(n\).
It is obvious that \(c^* = n \cdot s_{\alpha, \text{in}}\). 
The value~\(n\) depends on the output size~\(s_{\alpha, \text{out}}\) and input size \(s_{\beta, \text{in}}\), it determines how many of the~\(g_\alpha\) groups will influence the result of each of the~\(g_\beta\) groups.
%%%%%%%%%%
\begin{equation}
    n = \left\lceil \frac{s_{\beta, \text{in}}}{s_{\alpha, \text{out}}} \right\rceil = \left\lceil \frac{f_\alpha}{g_\beta} \cdot \frac{g_\alpha}{f_\alpha} \right\rceil = \left\lceil \frac{g_\alpha}{g_\beta} \right\rceil
\end{equation}
%%%%%%%%%%
We express the~CLC of a Split Convolutional Block~\((F_\alpha, F_\beta)\) as the fraction of input channels~\(c_\alpha\) that can impact each of the output channels~\(f_\beta\).
\begin{equation}
    \text{CLC}(F_\alpha, F_\beta) = \frac{c^*}{c_\alpha} = n \frac{s_{\alpha, \text{in}}}{c_\alpha} = \frac{1}{g_\alpha} \cdot \left\lceil \frac{g_\alpha}{g_\beta} \right\rceil
\end{equation}
As an example consider the split convolution visualized in Figure~\ref{fig:clc_example}. Using \(c_\alpha = 6, g_\alpha = 3, f_\alpha = c_\beta = 6, g_\beta = 2\), we obtain \(\text{CLC}(F_\alpha, F_\beta) = \left\lceil \frac{3}{2} \right\rceil \cdot \frac{1}{3} = \frac{2}{3}\). Four out of six input channels~\(c_\alpha\) will influence the result of each output channel.
%%%%%%
%%%%%%
A~CLC of~\(\frac{1}{3}\) means that the Split Convolutional Block realizes two completely separate data paths, while~\(\frac{2}{3}\) means that there is an overlap where some input groups impact the result of more than one output group.
%%%%%%%%%%%%%%%%%%%%%%%%%%%%%%%%%%%%%%%%%%%%%%%%%
\subsubsection{Split Configuration Choice}
The intuition is that high fan~in and high~CLC should provide higher network performance.
At the same time, we only want to consider configurations with high LUT-cost if their increased cost results in better performance.
Our proposed score~\(\mathcal{S}\) for split configurations is
\begin{equation}
 \mathcal{S}(F_\alpha, F_\beta) = \frac{\text{CLC}(F_\alpha, F_\beta)^2 \cdot \phi_\alpha \cdot \phi_\beta} {\log(\mathcal{C}(\phi_\alpha) + \mathcal{C(\phi_\beta}))^2},
\end{equation}
where~\(F_{\alpha,\beta}\) denote the parameter tuples of the first and second convolution and~\(\phi_{\alpha,\beta}\) their fan~ins.

We propose to find a network topology by choosing the top scoring configurations with a cost, that fits the use~case, train them and select the architecture with best fit.

\subsection{Toolchain and Hardware Implementation}
We implement the approach as an extension of our \blinded{elastic-AI creator} tool\footnote{open source available at~\blinded{https://github.com/es-ude/elasticai.creator.git}}. 
It automatically identifies precomputable blocks by their enclosing binary activations using pattern matching on an intermediate representation~(IR).
Then it replaces them in the IR with precomputed blocks by computing the corresponding truth tables. The right side of Figure~\ref{fig:maxpool_reorder} shows three such blocks in the post-training phase and their resulting precomputed counterparts.

Finally, the tool converts the IR to a fully pipelined RTL-level VHDL specification of the model.
% Since none of the precomputed operators takes logical time to process its inputs, logical delays are only introduced for buffering data in shift registers.
% These shift registers are just big enough to provide enough data for their succeeding operator and fully pipelined.
The resulting implementation can then be verified with a previously stored data set, e.g., a subset of the training data, via simulation.
Furthermore, the implementation does not use proprietary libraries or language constructs and is therefore portable to FPGAs from other manufacturers.
\section{Experiments} \label{sec:eval}
This section describes the data set, training of proposed model architecture, their results and hardware utilization.
%%%%%%%%%%%%%%%%%%%%%%%%%%%%%%%%%%%%%
\subsection{Dataset Handling and Training Properties}
We evaluate the performance of the split convolutions on the MIT-BIH atrial fibrillation dataset~\cite{moody1983new} with the binary classification problem of detecting atrial fibrillation in ECG~recordings.
The dataset contains~25~long-term ECG recordings, each with a duration of~10~hours and~two~channels.
The recordings have a resolution of~12~bits.
We subsample the records by selecting every second data point, resulting in a sampling rate of~\SI{125}{Hz}.
We partition~16~records into windows of about~\SI{42}{s}, resulting in about~13,800~examples, labelled as atrial fibrillation or sinus rhythm.
The first channel turned out to be sufficient for the problem, so we ignore the second channel.
For each split configuration, we perform an experiment consisting of~six~training runs over~400~epochs with a batch~size of~1,024. 
We use  binary cross entropy loss, the Adam optimizer with a learning rate of~\(5e^{-3}\) and decay the learning rate by~\(0.5\) every~50~epochs.
One run takes about~13~minutes using an NVIDIA RTX 8000 GPUs.
% \begin{figure}[htpb]
%     \includegraphics[width=0.95\linewidth]{images/model_architecture_mock_fix.pdf}
%     \label{fig:model_architecture}
% \caption{The prototype architecture for the atrial fibrillation use case.}
% \end{figure}
%%%%%%%%%%%%%%%%%%
\subsection{Model architecture and results}
\begin{figure}[tb] \centering
  \begin{tikzpicture}
    \begin{axis}[
        ymajorgrids,
        ymin=0.87, ymax=0.95,
        ytick distance=0.01,
        ylabel={Accuracy},
        xtick={1,2},
        xticklabels={Binarize -- Pooling, Pooling -- Binarize},
        width=8.2cm,
        height=5.5cm,
        grid=major,
        tick align=outside,
        boxplot/draw direction=y
    ]    
    % --- Binarize - Pooling (blau) ---
    \addplot+[
        boxplot prepared={
            median=0.883,
            upper quartile=0.885,
            lower quartile=0.879,
            upper whisker=0.890,
            lower whisker=0.874
        },
        fill=blue!30,
        draw=blue
    ] coordinates {
        (1,0.91) % Ausreißer
    };
    % --- Pooling - Binarize (orange) ---
    \addplot+[
        boxplot prepared={
            median=0.940,
            upper quartile=0.943,
            lower quartile=0.93,
            upper whisker=0.948,
            lower whisker=0.923
        },
        fill=orange!40,
        draw=orange
    ] coordinates {};
    \end{axis}
    \end{tikzpicture}
    \caption{Accuracy for different pooling positions for training.}
    \label{fig:ordering_accuracy}
    \vspace{-3mm}
\end{figure}
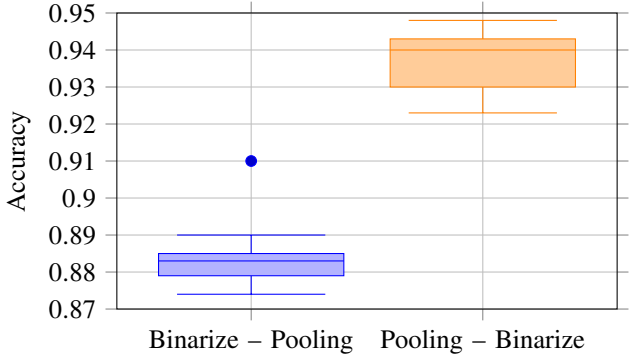

Table~\ref{tab:full_ecg_model} shows the architecture of the model based on previous work~\cite{einhausPrecomputed1DConvolutionalLayers2021}.
% After the first two convolutions and a max pooling layer, we place four convolutional blocks, each including batch norm, max pooling and a prelu activation function, denoted in the table as \textit{default block}.
Except for the first Split Convolutional Block, we use the same split configuration over the whole network, selected using Algorithm \ref{alg:split}.
% When splitting convolutions in default blocks, we use the same split configuration over the whole network.
%%%
\begin{table}[htpb] \small
  \caption{Network Architecture for MIT-BIH data set. 
  \(c_0\) denotes the number of input channels chosen. Experiments have been performed for \(c_0\) ranging from~6 to~12.} 
  \label{tab:full_ecg_model}
  
  % \begin{subtable}{\linewidth} \centering
    % \caption{\(c_0\) denotes the number of input channels chosen. Experiments have been performed for \(c_0\) ranging from~6 to~12.
    % \label{tab:ecg_model}}
    \rowcolors{2}{gray!15}{white}
    \begin{tabular}{lllll}
      \toprule
      Layer         & kernel size & in channels & out channels & stride \\\midrule
      conv1d        & 1           & 1           & 12           & 1      \\
      bnorm         & -           & -           & -            & -      \\
      binarize      & -           & -           & -            & -      \\
      SplitConv        & 10          & 12          & \(c_0\)      & 1      \\
      maxpool1d     & 8           & -           & -            & 6      \\
            SplitConv     & 6       & \(c_0\)  & \(c_0\)   & 1      \\
      maxpool1d     & 3       & -   & - & 2      \\
           SplitConv     & 6       & \(c_0\)  & \(c_0\)   & 1      \\
      maxpool1d     & 3       & -   & - & 2      \\
           SplitConv     & 6       & \(c_0\)  & \(c_0\)   & 1      \\
      maxpool1d     & 3       & -   & - & 2      \\
      % default block & 6           & \(c_0\)     & \(c_0\)      & 1      \\
      % default block & 6           & \(c_0\)     & \(c_0\)      & 1      \\
      % default block & 6           & \(c_0\)     & \(c_0\)      & 1      \\
      % default block & 6           & \(c_0\)     & \(c_0\)      & 1      \\
      linear        & 1           & \(c_0\)     & 1            & -      \\
      sigmoid       & -           & -           & -            & -      \\\bottomrule
    \end{tabular}
  %     \vspace*{0.5em}
  % \end{subtable}

  % \begin{subtable}{\linewidth}  \centering
  %   \caption{Structure of each default block with in/out channels from above.}
  %   \label{tab:default_block_ecg}
  %   \rowcolors{2}{gray!15}{white}
  %   \begin{tabular}{lll}
  %     \toprule
  %     Layer         & kernel size & stride \\\midrule
  %     SplitConv     & 6           & 1      \\
  %     maxpool1d     & 3           & 2      \\
  %     \bottomrule
  %   \end{tabular}
  % \end{subtable}
  % \vspace*{-1em}
\end{table}
%%%
To see the impact of the reordering step described in Section~\ref{sec:ordering}, we use this model architecture with~10~channels and train the two variants, (i) max pooling layer between batch norm and convolution, (ii) max pooling layer after binarization. Figure \ref{fig:ordering_accuracy} shows that the order has a signficant impact. The version with the pooling after binarization has a lower accuracy by~\SI{5}{\%}.
%%%%

Next, we want to understand the impact of our choice to always end Split Convolutional Blocks with a point-wise convolution (see Section~\ref{sec:split_convs}). 
The order of kernel sizes in a Split Convolutional Block has only little influence on the cost. 
Therefore, we study the resulting accuracy of both variants. 
The mean and standard deviation of accuracy achieved by the two kernel orders are~\(83.27 \pm 0.9 \%\) for~\((k_0, 1)\) and~\(81.05 \pm 0.7 \%\) for~\((1, k_0)\).
This indicates that kernel size order~\((k_0, 1)\) achieves better accuracy. This confirms our design choice. 
%%%
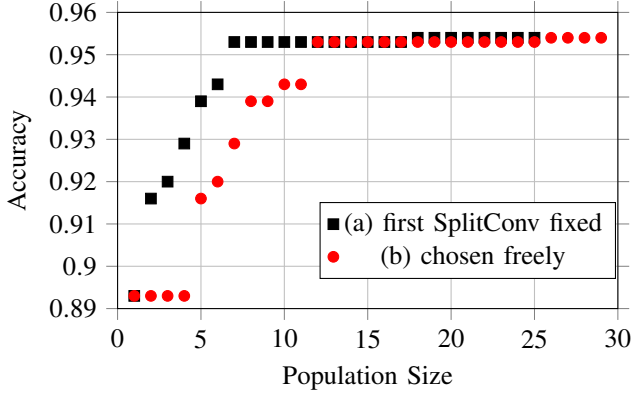
\begin{figure}[tb]    \centering
    \begin{tikzpicture}
    \begin{axis}[
        width=8.2cm,
        height=5.5cm,
        xlabel={Population Size},
        ylabel={Accuracy},
        xmin=0, xmax=30,
        ymin=0.89, ymax=0.96,
        ytick distance=0.01,
        grid=major,
        tick align=outside,
        enlargelimits=false,
        legend style={
            at={(0.7,0.1)},
            anchor=south,
        }
    ]
    \addplot[
        only marks,
        mark=square*,
        color=black
    ] coordinates {
        (1,0.893)
        (2,0.916)
        (3,0.920)
        (4,0.929)
        (5,0.939)
        (6,0.943)
        (7,0.953)
        (8,0.953)
        (9,0.953)
        (10,0.953)
        (11,0.953)
        (12,0.953)
        (13,0.953)
        (14,0.953)
        (15,0.953)
        (16,0.953)
        (17,0.953)
        (18,0.954)
        (19,0.954)
        (20,0.954)
        (21,0.954)
        (22,0.954)
        (23,0.954)
        (24,0.954)
        (25,0.954)
    };
    \addlegendentry{(a) first SplitConv fixed}
    \addplot[
        only marks,
        mark=*,
        color=red
    ] coordinates {
        (1,0.893)
        (2,0.893)
        (3,0.893)
        (4,0.893)
        (5,0.916)
        (6,0.920)
        (7,0.929)
        (8,0.939)
        (9,0.939)
        (10,0.943)
        (11,0.943)
        (12,0.953)
        (13,0.953)
        (14,0.953)
        (15,0.953)
        (16,0.953)
        (17,0.953)
        (18,0.953)
        (19,0.953)
        (20,0.953)
        (21,0.953)
        (22,0.953)
        (23,0.953)
        (24,0.953)
        (25,0.953)
        (26,0.954)
        (27,0.954)
        (28,0.954)
        (29,0.954)
    };
     \addlegendentry{(b) chosen freely}
    \end{axis}
    \end{tikzpicture}
    \caption{Accuracy of the best model in the population for increasing population sizes. 
    % The population is built from configurations with the highest score of all models that have a kernel sequence order of \((k_0, 1)\) with varying channels \(c_\alpha\). In (a) the first layer is fixed with \((c_\alpha, k_\alpha, g_\alpha, f_\alpha, k_\beta, g_\beta, f_\beta) = (12, 10, 12, 12, 1, 1, 12)\). In (b) the first layer is chosen freely. Scores are computed as means across all five layers.
    }
    \label{fig:acc_by_pop}
    \vspace{-5mm}
\end{figure}

Next, we want to find out how suitable our score is to find good split configurations.
For the first experiments, we keep the configuration for the first Split Convolutional Block fixed to correspond to a depthwise separable convolution
%, that is
%\[(c_\alpha=12, k_\alpha=10, g_\alpha=12, f_\alpha=12, k_\beta=1, g_\beta=12, f_\beta=12)\]
and only change the configurations for the other blocks.
This results in~73~possible split configurations.
To estimate the effort of finding a configuration with good accuracy that still fits on small FPGAs with 8k~LUTs.
We sort these configurations by their score, choose an increasing population of the first~\(n\) configurations (varied from~1 to~30), train them, and select the model with the best accuracy.
The result is visualized with case~(a) in Figure~\ref{fig:acc_by_pop}. The approach reaches a plateau after increasing the population size to~seven.

Typically, one needs to find configurations for many different layers, that are varied independently of each other.
Thus, we repeat the same experiment but also vary the configuration for the first Split Convolutional Block.
To set a score for the whole network, we compute the mean over the scores of all layers.
Increasing the population size again, we find the result visualized with case (b) in Figure~\ref{fig:acc_by_pop}.
Due to the filtering for analytic LUT-costs below~8,000, this results in 115 possible split configurations.
With this setup, the approach reaches the plateau for a population size of~twelve.
The plateau is reached in both cases after increasing the population size to contain about~\SI{10}{\%} of the full configuration set resulting in a drastic search space reduction.

Next, we evaluate how well the score captures the intuition of rewarding models for a good trade-off between cost and accuracy in comparison to other candidates.
We can formalize the intuition as follows: Let \(i\), \(j\) be any two of these split configurations, \(A_i, A_j\) their corresponding accuracy, \(\mathcal{S}_i, \mathcal{S}_j\) their score and \(\mathcal{C}_i, \mathcal{C}_j\) their analytic cost, then we should always find
%%%
\begin{equation}
  \mathcal{S}_i < \mathcal{S}_j \Rightarrow (A_i < A_j) \lor (\mathcal{C}_i > \mathcal{C}_j) \label{eq:score_condition}
\end{equation}
%%%
Again, we keep the first Split Convolutional Block fixed with a configuration that corresponds to a depth-wise separable convolution and only vary the other blocks.
We choose a channel size of twelve, as this allows for more split configurations than values between six and eleven.
We generate split configurations using Algorithm~\ref{alg:split} with a maximum fan~in \(\phi_\text{max}= 12\).
Removing all configurations that would increase the LUT-cost to more than~8,000~leaves~24~possible choices for the other blocks.
%%%
For the~24~remaining choices we find that~275~out of~276~pairs fulfil Equation~\eqref{eq:score_condition}.
The single outlier pair can be found in the last row of Table~\ref{tab:counter_examples_intuition}.
Both configurations have a very small score, high cost and low accuracy compared to pareto-optimal configurations: Thus, both models would not be considered for training in a practical scenario.

When allowing the number of input channels for each channel to be chosen freely, we obtain~73~possible choices, resulting in~2,628~pairs.
Table~\ref{tab:counter_examples_intuition} lists all pairs that violate Equation~\eqref{eq:score_condition}.
Only two of these pairs show an accuracy above~\SI{90}{\%} with a difference of~\SI{1}{\%}.
%%%%
\begin{table}[htpb]
\centering
\caption{Pairs that violate \eqref{eq:score_condition}. 
%Every third row, lists the difference between the configuration with the higher score~\(j\) and the configuration~\(i\) that outperforms it wrt. accuracy and analytic LUT cost. T
The first Split Convolutional Block is fixed to correspond to a depthwise separable configuration.
%while all other layers have a kernel sequence of~\((k_0, 1)\). All other parameters vary. Split Configuration specified as a tuple \((c_\alpha, k_\alpha, g_\alpha, f_\alpha, k_\beta, g_\beta, f_\beta)\).
\label{tab:counter_examples_intuition}}
\begin{tabular}{lrrrrr}
\toprule
&   Split Config.  & Score & LUTs & Acc. & F1 \\
\midrule
\(j\) & \((10, 6, 10, 10, 1, 1, 10)\) & 20.62 & 3,087 & 93.86 & 93.31\\
\(i\) & \((12, 6, 12, 24, 1, 3, 12)\) & 6.52 & 2,713 & 93.92 & 93.41\\
% diff. &&& 374 & -0.06 & -0.1\\

\rowcolor{gray!15} \(j\) & \((10, 6, 10, 20, 1, 2, 10)\) & 10.14 & 3,127 & 93.03 & 92.49 \\
\rowcolor{gray!15} \(i\) & \((12, 6, 12, 24, 1, 3, 12)\) & 6.52 & 2,713 & 93.92 & 93.41\\ 
% \rowcolor{gray!15} diff. &&& 414 & -0.89 & -0.92\\

\(j\) & \((6, 6, 6, 24, 1, 6, 6)\) & 1.07 & 2,059 & 75.61 & 75.09\\
\(i\) & \((6, 6, 6, 18, 1, 6, 6)\) & 0.70 & 2,011 & 76.51 & 75.08\\
% diff. &&& 48 & -0.91 & +0.01\\

\rowcolor{gray!15} \(j\) & \((8, 6, 8, 32, 1, 8, 8)\) & 0.69 & 2,293 & 76.10 & 75.17\\
\rowcolor{gray!15} \(i\) & \((7, 6, 7, 21, 1, 7, 7)\) & 0.55 & 2,120 & 76.38 & 75.01\\
% \rowcolor{gray!15} diff. &&& 173 & -0.27 & +0.16\\

\(j\) & \((8, 6, 8, 8, 1, 4, 8)\) & 0.59 & 2,133 & 74.35 & 72.11\\
\(i\) & \((7, 6, 7, 21, 1, 7, 7)\) & 0.55 & 2,120 & 76.38 & 75.01\\
% diff. &&& 13 & -2.03 & -2.90\\

\rowcolor{gray!15} \(j\) & \((8, 6, 8, 32, 1, 8, 8)\) & 0.69 & 2,293 & 76.10 & 75.17\\
\rowcolor{gray!15} \(i\)& \((8, 6, 8, 24, 1, 8, 8)\) & 0.45 & 2,229 & 76.60 & 74.92\\
% \rowcolor{gray!15} diff. &&& 64 & -0.50 & -0.25\\

\(j\) &\((10, 6, 10, 10, 1, 5, 10)\) & 0.41 & 2,327 & 74.65 & 74.19 \\
\(i\) &\((8, 6, 8, 16, 1, 8, 8)\) & 0.25 & 2,165 & 74.79 & 72.27\\
% diff. && &162 & -0.14  & +1.92\\

\rowcolor{gray!15}\(j\) & \((12, 6, 6, 12, 1, 12, 12)\) & 0.08 & 6,505 & 73.21 & 71.16 \\
\rowcolor{gray!15}\(i\) & \((12, 6, 6, 6, 1, 6, 12)\) & 0.05 & 4,465 & 75.50 & 72.89\\
% \rowcolor{gray!15}diff. & && 2,040 & -2.28  & -1.73\\
\bottomrule
\end{tabular}
\end{table}
%%%
Next, we want to find out how many of the best architectures wrt. the score must be trained to find the models that make up the pareto front.
Ideally, the supremum of these architectures would correspond to that front.
%%%
\begin{table}[htbp] \centering
\caption{Pareto front of split configurations with depthwise separable first Split Convolutional Block.
% when keeping the first split configuration fixed at~\((c_\alpha, k_\alpha, g_\alpha, f_\alpha, k_\beta, g_\beta, f_\beta) = (12, 10, 12, 12, 1, 1, 12)\) and varying the~\(c_\alpha\) for the default blocks. The accuracy is computed as the mean over six trials.
}
\label{tab:pareto_default_blocks_ecg} 
\begin{tabular}{lrrrr}
  \toprule
  Split Configuration  & &&& \\
  \((c_\alpha, k_\alpha, g_\alpha, f_\alpha, k_\beta, g_\beta, f_\beta)\) & \multirow{-2}{*}{Score} & \multirow{-2}{*}{LUTs}& \multirow{-2}{*}{Acc.} & \multirow{-2}{*}{F1}  \\
  \midrule
  \((12,6,12,36,1,3,12)\) & 5.94 & 6,601& 95.37 & 94.95  \\
\rowcolor{gray!15}\((12,6,12,12,1,1,12)\) & 17.94 & 6,505& 95.34 & 94.94  \\
\((12,6,6,6,1,1,12)\) & 11.03& 4,465 & 94.40 & 93.93  \\
\rowcolor{gray!15}\((11,6,11,11,1,1,11)\) & 19.00  & 4,228& 94.31 &93.83 \\
\((12,6,12,24,1,3,12)\) & 6.52 & 2,713& 93.92& 92.29  \\
\rowcolor{gray!15}\((9,6,9,9,1,1,9)\) & 22.17 & 2,554& 92.93 & 92.30  \\
\((8,6,8,16,1,2,8)\) & 11.85& 2,261  & 92.40 & 91.81\\
\rowcolor{gray!15}\((8,6,8,8,1,1,8)\) & 25.62 & 2,229& 92.05 &91.41  \\
\((7,6,7,7,1,1,7)\) & 26.48 & 2,064& 91.63 &91.10 \\
\rowcolor{gray!15}\((6,6,6,12,1,2,6)\) & 12.93& 1,939 & 89.51 &88.49 \\
\((6,6,6,6,1,1,6)\) & 34.98  & 1,915 & 89.30 & 88.47\\
  \bottomrule
\end{tabular}
\vspace*{-1.6em}
\end{table}
%%%
Table~\ref{tab:pareto_default_blocks_ecg} shows this pareto front.
The first thing to note is that one of the eleven configurations also appears in Table~\ref{tab:counter_examples_intuition} as it violates the score condition. 
The configuration~\((12,6,12,24,1,3,12)\) in the fifth row of Table~\ref{tab:pareto_default_blocks_ecg} is part of the pareto front and outperforms two other configurations with higher score.
While the accuracy of the configuration on the first row of Table~\ref{tab:pareto_default_blocks_ecg} is only slightly better than that of the configuration below with higher score, looking at the value for~\(f_\alpha\) indicates that the score underestimates this value's impact.
To cover the full pareto front it is sufficient to apply a threshold of~\(>5.0\) to the set of configurations. This would select~19~out of~73~possible configurations.
Put differently, training~\SI{26}{\%} of the best configurations wrt. the score is sufficient to cover the pareto front in this experiment.

\subsection{Hardware Utilization}
All hardware accelerators are synthesized with AMD~Vivado~2023.1 and deployed on the AMD~Spartan-7~S15 FPGA. 

In addition to the neural network, the accelerator also features a BRAM input buffer to store an entire data sample and logic that allows control via an SPI interface.
The required clock cycles are mainly dictated by the size of the data sample, as data is processed in a sequentially to save area.
The system takes one clock cycle per time step of the data sample.
To avoid measuring the~SPI~communication between~MCU and~FPGA, we modify the hardware implementation to perform~thousand~inferences instead of one before signalling done.
For an atrial fibrillation model with depthwise-separable styled split convolutions, we measure the time starting between signalling start of computation until the accelerator signals done.
We measure~\SI{50.88}{\micro s} per inference at a clock period of~\SI{10}{ns}, i.e., approximately~5,088~clock cycles per inference, close to the~5,085~clock cycles we read from simulation.
We can successfully generate the design for clock rates of up to~\SI{143}{MHz}.

By the nature of our approach, none of the synthesized designs consumes any DSP slices.
The previous analytic cost estimations only consider the implementation of the neural network and ignore any of the logic required to interact with the MCU or manage the BRAM input and output buffers.
For all our synthesized designs Vivado estimates a total power consumption between~\SI{42}{mW} and~\SI{65}{mW} with~\SI{20}{mW} of static power consumption. The numbers show a weak negative correlation of~{-0.4} between analytic LUT cost and power consumption.
The power consumption estimated for the network component directly correlates with the networks size and ranges from~\SI{2}{mW} (network with~200~LUTs) to~\SI{5}{mW} (networks with~8k~LUTs).
As the reported post optimization LUT costs include additional components, the actual LUT consumption for very small networks with less than~600~analytic LUTs tends to be almost doubled.
As networks grow larger the static overhead from other logic plays a smaller role and the use of dedicated muxing primitives and other logic optimizations become more important. As a result, bigger models (for which we analytically estimate three to 6k~LUTs) need only half as many LUTs.
%%%%%%%%%%%%%%%%%%%%%%%%%%%%%%%%%%%%%
\subsection{Comparison to related work}
Finally, we compare our performance with related work on atrial fibrillation on FPGAs.  Table~\ref{tab:comparison} summarizes the hardware utilization (LUTs, DSP slices, FFs) and the network performance (accuracy, F1, latency) for different model architectures. All examples are detecting atrial fibrillation using the MIT-BIH database. We compare with two of our selected configurations. BIG is the best selected model that still fits on our target platform. It has \((12, 10, 12, 12, 1, 1, 12)\) for its first Split Convolutional Block and \((12, 6, 12, 12, 1, 1, 12)\) for all others. SMALL for comparison is our best tiny network with \((12, 10, 12, 12, 12, 2, 10)\) for its first Split Convolution and \((10, 6, 10, 10, 1, 2, 10)\) for its other blocks. 
%%%%
\begin{table}[ht] \centering

\caption{Comparison to other atrial fibrillation detectors using different model architectures on hardware}
\begin{tabular}{l|ccccc}
\toprule
Approach                                                    & LUTs          & DSP           & FF        & Accuracy              & Latency \\
                                                            &               &               &           & / F1                    & [\textmu s] \\
\midrule
ViT \cite{chandrasekaran2025_Cardiologist}\(^\chi\)         & 53,810        & 131           & 24,026    & 99.0 / \textbf{99.4}  & 3,830  \\

\rowcolor{gray!15} 1D Incept. \cite{Nhat2025_EfficientBeat}\(^\chi\)           & 26,987        & 448           & -         & 99.3 / 99.0           & 231 \\
\rowcolor{gray!15} 1D Incept. \cite{pham2025_FastEffEcg}\(^\chi\)              & 28,758        & 80            & 13,313    & \textbf{99.5} / 99.2  & \textbf{22} \\
2D-CNN \cite{greco2025_FastArrhythmia}\(^{\chi, \beta}\)    & 20,198        & 98            & 12,993    & 91.9 / -              & - \\
\rowcolor{gray!15} 1D-CNN \cite{tiany2024_DLAFDet}\(^\chi\)                    & 8,752         & 77            & 5,303     & 96.4 / 93.5           & 66 \\
\rowcolor{gray!15} 1D-CNN \cite{loroch2025_lowpowercardio}\(\gamma\)           & 3,802         & 5             & 1,848     & - /95.8               & - \\
arrWNN\cite{pillai2024_arrWNN}$^{\alpha, \rho}$             & -             & \textbf{0}    & -         & 88.2 / 81.3           & -\\
\rowcolor{gray!15} BIG (ours)$^{\chi, \rho}$                           & 2,844         & \textbf{0}    & 871      & 95.8 / 95.6           & 51 \\
\rowcolor{gray!15} 1D-CNN\cite{einhausPrecomputed1DConvolutionalLayers2021}$^{\chi,\rho}$   & 2,352  & \textbf{0} & -      & 94.2 / -              & 52\\

1D-CNN\cite{einhausPrecomputed1DConvolutionalLayers2021}$^{\chi, \rho}$  & 986 & \textbf{0} & -         & 82.4 / -              & 52\\

SMALL (ours)$^{\chi, \rho}$ 
& \textbf{536}  & \textbf{0}    & \textbf{785} & 88.4 / 87.4        & 51 \\

\bottomrule
\multicolumn{6}{l}{\footnotesize{$\alpha$: custom IC, $\gamma$: Lattice iCE40UP5K, $\chi$: AMD FPGAs}}\\
\multicolumn{6}{l}{$\beta$: four labels, $\rho$: precomputed, $-$: not specified}
\end{tabular}
\label{tab:comparison}
\end{table}

% The vision transformer (ViT)~\cite{chandrasekaran2025_Cardiologist} achieves the highest F1-score but it has also the highest utilization and latency. This implementation is not suitable for deploying on small FPGAs.

The results show that our models achieve competitive performance with very tight hardware constraints. SMALL has the smallest footprint and acceptable performance. BIG performs comparable to \cite{loroch2025_lowpowercardio} with a far smaller footprint. 

The Vision Transformer ViT~\cite{chandrasekaran2025_Cardiologist} and the 1D-CNNs with inception blocks achieve the highest accuracy and F1-score, but all of them drastically exceed out target platform's resources.
The 2D-CNN is even outperformed by the non-precomputed 1D-CNNs wrt. resource requirements and performance.
The only non LUT-based approach suitable for our hard resource constraints is the implementation of a 1D-CNN found through HW-aware NAS by Loroch et al.~\cite{loroch2025_lowpowercardio}, using a Lattice instead of an AMD~FPGA.
The last five rows show LUT-based approaches.
The arrWNN~\cite{pillai2024_arrWNN} uses a custom IC instead of an FPGA. Their resource utilization cannot be compared directly.

% Compared to previous work based on depthwise separable convolutions~\cite{einhausPrecomputed1DConvolutionalLayers2021} we reduce resource consumption by a factor~2 for a bigger model with twelve channels, or~1.5 for smaller model with six channels, while improving network performance.
% \section{HW Design}
% \input{text/evaluation}
% \section{evaluation}
\section{Conclusion}\label{sec:conclusion}
We reduce the fan-in of a 1D convolutional neural network to make LUT-based precomputation for ultra-low resource FPGAs feasible. This enables the realization of smart embedded sensor systems that use local machine learning to process sensor data.
We defined a score based on fan~in, LUT-costs and layer connectivity that allows us to estimate the models' trade-off between performance and costs. With this we show how to reduce the search space of possible architectures while staying pareto optimal.
We obtain very efficient accelerators with good performance compared to related work.
In future work, we want to evaluate our approach on additional time-series datasets, e.g., the CHB-MIT database for seizure prediction~\cite{guttag_chb-mit_2010}, to further study generalizability. Additionally, we want to explore combining our approach with others such as NeuraLUT~\cite{andronic2025neuralut} or logic gate networks\cite{petersen2022_LogicGateNetwork}.

\bibliographystyle{IEEEtran}
\bibliography{IEEEabrv,zotero_refs}

@inproceedings{andronic2023polylut,
  title={{PolyLUT: learning piecewise polynomials for ultra-low latency FPGA LUT-based inference}},
  author={Andronic, Marta and Constantinides, George A},
  booktitle={IEEE ICFPT},
  pages={60--68},
  year={2023},
}

@article{mattioli20221d,
  title={{A 1D CNN for high accuracy classification and transfer learning in motor imagery EEG-based brain-computer interface}},
  author={Mattioli, Francesco and Porcaro, Camillo and Baldassarre, Gianluca},
  journal={J. Neu. Eng.},
  volume={18},
  number={6},
  year={2022},
  publisher={IOP Publishing}
}

@inproceedings{li2017classification,
  title={{Classification of ECG signals based on 1D convolution neural network}},
  author={Li, Dan and Zhang, Jianxin and Zhang, Qiang and Wei, Xiaopeng},
  booktitle={IEEE Healthcom},
  pages={1--6},
  year={2017}
}

@article{eren2019generic,
  title={{A generic intelligent bearing fault diagnosis system using compact adaptive 1D CNN classifier}},
  author={Eren, Levent and Ince, Turker and Kiranyaz, Serkan},
  journal={Jour. Sig. Pro. sys.},
  volume={91},
  number={2},
  pages={179--189},
  year={2019},
  publisher={Springer}
}

@article{ince2016real,
  title={{Real-time motor fault detection by 1-D convolutional neural networks}},
  author={Ince, Turker and Kiranyaz, Serkan and Eren, Levent and Askar, Murat and Gabbouj, Moncef},
  journal={IEEE Trans. Ind. Electron.},
  volume={63},
  number={11},
  pages={7067--7075},
  year={2016},
  publisher={IEEE}
}

@inproceedings{avci2017structural,
  title={{Structural damage detection in real time: implementation of 1D convolutional neural networks for SHM applications}},
  author={Avci, Onur and Abdeljaber, Osama and Kiranyaz, Serkan and Inman, Daniel},
  booktitle={IMAC},
  pages={49--54},
  year={2017}
}

@article{kiranyaz2015real,
  title={{Real-time patient-specific ECG classification by 1-D convolutional neural networks}},
  author={Kiranyaz, Serkan and Ince, Turker and Gabbouj, Moncef},
  journal={IEEE Trans. Biom. Eng.},
  volume={63},
  number={3},
  pages={664--675},
  year={2015},
  publisher={IEEE}
}

@article{mittal2020survey,
  title={{A survey of FPGA-based accelerators for convolutional neural networks}},
  author={Mittal, Sparsh},
  journal={Neu. comp. and appli.},
  volume={32},
  number={4},
  year={2020},
  publisher={Springer}
}

@article{courbariauxBinarizedNeuralNetworks2016,
  title = {{Binarized {{Neural Networks}}: {{Training Deep Neural Networks}} with {{Weights}} and {{Activations Constrained}} to +1 or -1}},
  author = {Hubara, Itay and Courbariaux, Matthieu and Soudry, Daniel and {El-Yaniv}, Ran and Bengio, Yoshua},
  year = {2016},
  month = feb,
  journal = {NeurIPS}
}

@inproceedings{einhausPrecomputed1DConvolutionalLayers2021,
  title = {{Towards {{Precomputed 1D-Convolutional Layers}} for {{Embedded FPGAs}}}},
  booktitle = {ECML PKDD},
  author = {Einhaus, Lukas and Qian, Chao and Ringhofer, Christopher and Schiele, Gregor},
  year = {2021},
  pages = {327--338},
}

@inproceedings{andronic2024neuralut,
  title={{NeuraLUT: Hiding neural network density in boolean synthesizable functions}},
  author={Andronic, Marta and Constantinides, George A},
  booktitle={IEEE FPL},
  pages={140--148},
  year={2024},
}

@inproceedings{andronic2025neuralut,
  title={{NeuraLUT-Assemble: Hardware-aware Assembling of Sub-Neural Networks for Efficient LUT Inference}},
  author={Andronic, Marta and Constantinides, George A},
  booktitle={IEEE FCCM},
  pages={208--216},
  year={2025}
}

@article{nazemiNullaNetTinyUltralowlatency2021,
  title = {{{NullaNet Tiny}}: {{Ultra-low-latency DNN Inference Through Fixed-function Combinational Logic}}},
  author = {Nazemi, Mahdi and Fayyazi, Arash and Esmaili, Amirhossein and Khare, Atharva and Shahsavani, Soheil Nazar and Pedram, Massoud},
  year = {2021},
  month = apr,
  journal = {arXiv:2104.05421},
}

@inproceedings{rastegariXNORnetImagenetClassification2016,
  title = {{{XNOR-net}}: {{Imagenet}} Classification Using Binary Convolutional Neural Networks},
  booktitle = {Lecture {{Notes}} in {{Computer Science}}},
  author = {Rastegari, Mohammad and Ordonez, Vicente and Redmon, Joseph and Farhadi, Ali},
  year = {2016},
  volume = {9908 LNCS},
}

@article{umurogluLogicNetsCoDesignedNeural2020,
  title = {{{LogicNets}}: {{Co-Designed Neural Networks}} and {{Circuits}} for {{Extreme-Throughput Applications}}},
  author = {Umuroglu, Yaman and Akhauri, Yash and Fraser, Nicholas J. and Blott, Michaela},
  year = {2020},
  month = apr,
  journal = {arXiv:2004.03021},
}

@article{moody1983new,
  title={{A new method for detecting atrial fibrillation using RR intervals}},
  author={Moody, George},
  journal={Proc. Comput. Cardiol.},
  volume={10},
  pages={227--230},
  year={1983}
}

@misc{guttag_chb-mit_2010,
	title = {{CHB}-{MIT} {Scalp} {EEG} {Database} (version 1.0.0)},
	url = {https://physionet.org/content/chbmit/1.0.0/},
	doi = {10.13026/C2K01R},
	urldate = {2024-01-01},
	publisher = {physionet.org},
	author = {Guttag, John},
	year = {2010},
}

@inproceedings{ebrahimiIterativePruningAlgorithm2023,
  title = {{Iterative Pruning Algorithm for Efficient Look-up Table Implementation of Binary Neural Networks}},
  booktitle = {{{IEEE NEWCAS}}},
  author = {Ebrahimi, Amirali and Pullu, Vineeth Narayan and Pierre Langlois, J.M. and David, Jean-Pierre},
  year = {2023},
  pages = {1--5},
}

@inproceedings{zhang2018shufflenet,
  title={{Shufflenet: An extremely efficient convolutional neural network for mobile devices}},
  author={Zhang, Xiangyu and Zhou, Xinyu and Lin, Mengxiao and Sun, Jian},
  booktitle={IEEE CVPR},
  pages={6848--6856},
  year={2018}
}

@inproceedings{Loehler2024,
  author       = {Philipp Löhler and Andreas Albert and Laura Heyermann and Gregor Schiele and Karsten Seidl and Andreas Erbslöh},
  title        = {{Classification of ON‐ and OFF‐Retinal Ganglion Cell Types in Extracellular Recordings}},
  booktitle    = {Proc. Workshop Biosignal},
  year         = {2024},
  doi          = {10.47952/gro-publ-209},
}

@INPROCEEDINGS{pillai2024_arrWNN,
  author={Pillai, Velu and Miranda, Igor D. S. and Musale, Tejas and Jadhao, Mugdha and Souza Neto, Paulo C. R. and Susskind, Zachary and Bacellar, Alan T. L. and Lhostis, Mael and Lima, Priscila M. V. and Dutra, Diego L. C. and John, Eugene B. and Breternitz, Mauricio and França, Felipe M. G. and Ozer, Emre and John, Lizy K.},
  booktitle={IEEE IFETC}, 
  title={{arrWNN: Arrhythmia-Detecting Weightless Neural Network FlexIC}}, 
  year={2024},
  pages={1-4}
}

@article{chandrasekaran2025_Cardiologist,
author = {Chandrasekaran, Saravanakumar and Chandran, Srinivasan and Selvam, Immaculate Joy},
title = {{FPGA-Based Implementation of Real-Time Cardiologist-Level Arrhythmia Detection and Classification in Electrocardiograms Using Novel Deep Learning}},
journal = {J. Circuit Theory and Applications},
volume = {53},
number = {6},
pages = {3662-3683},
doi = {https://doi.org/10.1002/cta.4289},
year = {2025}
}

@INPROCEEDINGS{tiany2024_DLAFDet,
  author={Tianyi, Huang and Yuchen, Jiang and Yijing, Wang and Qinghui, Lyu and Dakun, Lai},
  booktitle={IEEE PRML}, 
  title={{Deep Learning Based Automatic Detection Algorithm of Atrial Fibrillation Implemented on FPGA}}, 
  year={2024},
  pages={330-334},
  doi={10.1109/PRML62565.2024.10779948}
}

@INPROCEEDINGS{Nhat2025_EfficientBeat,
  author={Nhat, Nam Le Nguyen and Tran, Thi Diem},
  booktitle={IEEE ICDV}, 
  title={{Efficient ECG Beat Classification Using Inception Network on Software and FPGA Platforms}}, 
  year={2025},
  pages={67-72},
  doi={10.1109/ICDV66179.2025.11135057}
}

@INPROCEEDINGS{pham2025_FastEffEcg,
  author={Pham, Hoai Luan and Tran, Thi Diem and Le, Vu Trung Duong and Vu, Tuan Hai and Nakashima, Yasuhiko},
  booktitle={Int. Conf. ECTI-CON}, 
  title={{A Fast and Memory-Efficient CNN Accelerator for ECG Classification in Remote Healthcare Systems}}, 
  year={2025},
  volume={},
  number={},
  pages={1-6},
}

@article{greco2025_FastArrhythmia,
title = {{Fast and low cost FPGA-based architecture for arrhythmia detection with CNN}},
journal = {Internet of Things},
volume = {33},
pages = {101705},
year = {2025},
author = {Luca Greco and Francesco Moscato and Pierluigi Ritrovato and Mario Vento},
}

@INPROCEEDINGS{loroch2025_lowpowercardio,
  author={Loroch, Dominik and Feldmann, Johannes and Rybalkin, Vladimir and Wehn, Norbert},
  booktitle={IEEE SOCC}, 
  title={{Low-power, Energy-efficient, Cardiologist-level Atrial Fibrillation Detection for Wearable Devices}}, 
  year={2025},
  volume={},
  number={},
  pages={1-6},
  doi={10.1109/SOCC66126.2025.11235473}
}

@inproceedings{khatei2024_compressedLUT,
author = {Khataei, Alireza and Bazargan, Kia},
title = {{CompressedLUT: An Open Source Tool for Lossless Compression of Lookup Tables for Function Evaluation and Beyond}},
year = {2024},
doi = {10.1145/3626202.3637575},
booktitle = {ACM/SIGDA Int. Symp. on FPGAs},
pages = {2–11},
numpages = {10},
}

@inproceedings{cassidy2025_reducedLUT,
author = {Cassidy, Oliver and Andronic, Marta and Coward, Samuel and Constantinides, George A.},
title = {{ReducedLUT: Table Decomposition with "Don't Care" Conditions}},
year = {2025},
doi = {10.1145/3706628.3708823},
booktitle = {Proc. ACM/SIGDA Int. Symp. on FPGAs},
pages = {36–42},
numpages = {7}
}

@inproceedings{aleksander2009_introWNN,
    author = {Aleksander, I. and De Gregorio, M. and Franca, F.M.G and Lima, P.M.V. and Morton H.},
    title = {{A brief introduction to Weightless Neural Systems}},
    year = {2009},
    booktitle = {ESANN},
    isbn = {2930307099}
}

@inproceedings{petersen2024_LogicGateConv,
 author = {Petersen, Felix and Kuehne, Hilde and Borgelt, Christian and Welzel, Julian and Ermon, Stefano},
 booktitle = {NeurIPS},
 doi = {10.52202/079017-3851},
 pages = {121185--121203},
 title = {{Convolutional Differentiable Logic Gate Networks}},
 volume = {37},
 year = {2024}
}

@inproceedings{petersen2022_LogicGateNetwork,
 author = {Petersen, Felix and Borgelt, Christian and Kuehne, Hilde and Deussen, Oliver},
 booktitle = {NeurIPS},
 pages = {2006--2018},
 title = {{Deep Differentiable Logic Gate Networks}},
 volume = {35},
 year = {2022}
}

@inproceedings{Nag_2025,
   title={{LL-ViT: Edge Deployable Vision Transformers with Look Up Table Neurons}},
   DOI={10.1109/icfpt67023.2025.00013},
   booktitle={IEEE ICFPT},
   author={Nag, Shashank and Bacellar, Alan T.L. and Susskind, Zachary and Jha, Anshul and Liberty, Logan and Sivakumar, Aishwarya and John, Eugene B. and Kailas, Krishnan and Lima, Priscila M.V. and Yadwadkar, Neeraja J. and França, Felipe M.G. and John, Lizy K.},
   year={2025},
   pages={19–29}
}

@inproceedings{ling2025_strikewatch,
    title={{StrikeWatch: Wrist-worn Gait Recognition with Compact Time-series Models on Low-power FPGAs}},
    author={Ling, Tianheng and Qian, Chao and Zdankin, Peter and Weis, Torben and Schiele, Gregor},
    booktitle={IEEE AIoT},
    year={2025}
}

@inproceedings{krizhevsky2012_imagenet,
  title={{ImageNet Classification with Deep Convolutional Neural Networks}},
  author={A. Krizhevsky, I. Sutskever, G. Hinton},
  booktitle={NeurIPS},
  year={2012}
}
\end{document}